\documentclass[preprint,aps,prb,showpacs]{revtex4}
\begin{document}
\title{Electronic structure of and Quantum size effect in III-V and II-VI semiconducting nanocrystals
using a realistic tight binding approach.}
\author{Ranjani Viswanatha$^{1}$}
\author{Sameer Sapra$^{1}$}
\author{Tanusri Saha-Dasgupta$^{2}$}
\author{D. D. Sarma$^{1}$}
\altaffiliation[Also at ] {Jawaharlal Nehru Centre for Advanced
Scientific Research, Bangalore-560054,India and Centre for Condensed
Matter Theory, Indian Institute of Science.}
\email{sarma@sscu.iisc.ernet.in} \affiliation{$^{1}$ Solid State and
Structural Chemistry Unit, Indian Institute of Science, Bangalore
- 560012, India\\
$^{2}$ S.N. Bose Centre, Kolkatta, India. \\}

\begin{abstract}
We analyze the electronic structure of group III-V semiconductors
obtained within full potential linearized augmented plane wave
(FP-LAPW) method and arrive at a realistic and minimal tight-binding
model, parameterized to provide an accurate description of both
valence and conduction bands. It is shown that cation $sp^3-$ anion
$sp^3d^5$ basis along with the next nearest neighbor model for
hopping interactions is sufficient to describe the electronic
structure of these systems over a wide energy range, obviating the
use of any fictitious $s^*$ orbital, employed previously. Similar
analyses were also performed for the II-VI semiconductors, using the
more accurate FP-LAPW method compared to previous approaches, in
order to enhance reliability of the parameter values.  Using these
parameters, we calculate the electronic structure of III-V and II-VI
nanocrystals in real space with sizes ranging upto about 7~nm in
diameter, establishing a quantitatively accurate description of the
band-gap variation with sizes for the various nanocrystals by
comparing with available experimental results from the literature.
\end{abstract}

\pacs{71.20.Nr, 73.22.-f, 78.67.Bf}

\maketitle

\section{Introduction:}

Semiconductor nanocrystals, with the tunability of their electronic
and optical properties by the three-dimensional confinement of
carriers, have attracted considerable interest as technologically
important materials.~\cite{devices}  Hence, the study of the quantum
confinement in these semiconductors has been a subject of intense
study. Though the first approach to obtain a quantitative
understanding of the quantum confinement effects on the bandgap of
the nanocrystal as a function of size was given by the Effective
Mass Approximation~\cite{ema} (EMA), it is well known to
overestimate the bandgap in the lower size regime. In the past few
decades, theoretical predictability of the variation of bandgap as a
function of size has increased due to the development of a host of
different theoretical approaches, starting from the {\it ab-initio}
methods~\cite{abinitio} to the semi-empirical
pseudo-potential~\cite{ppmeth1,ppmeth2} and tight
binding~\cite{Liplan,Jancu,Allan,DD1,DDInP,DD2,DDZnO} (TB)
approaches. Recently, the TB method has gained certain popularity,
both because of its realistic description of structural and
dielectric properties in terms of chemical bonds and its simplicity
enabling one to handle very large systems. Slater-Koster's
suggestion to treat the TB model as an interpolation
scheme~\cite{SlatKost} has been widely used in various
semiconductors.  However, the intuitively appealing, nearest
neighbor $sp^3$ model fails to explain even the indirect gap in most
of the III-V semiconductors satisfactorily, especially at the
X-point. In order to mimic the influence of the excited $d$ states,
Vogl {\it et al.} used the $s^*$ orbital, in an {\it ad hoc}
manner.~\cite{vogl} Though it could explain the bandgap at the X
point correctly, the band curvatures were not properly described.
Following the recognition of the importance of the $d$ states by the
pseudo-potential methods, Jancu {\it et al.} have recently performed
a TB calculation using a $sp^3d^5s^*$ basis for both cations and
anions and the nearest neighbor interactions on III-V as well as
group IV semiconductors.~\cite{Jancu} The band dispersions obtained
by this calculation is found to overcome most of the deficiencies of
the earlier TB models, though the use of the $s^*$-orbital,
originally included to account for the absence of the excited $d$
states, becomes more questionable with the inclusion of the $d$
orbitals in the basis.   Moreover, the transferability of the TB
parameters obtained from bulk {\it ab-initio} band structures to the
nanometric regime remains a controversial issue.

With the advance of experimental techniques such as photoemission
and inverse photoemission techniques, it is possible to map out the
density of states (DOS) of both valence and conduction bands for the
bulk materials as well as the nanocrystals.  With the introduction
of site and angular-momentum specific X-ray emission and absorption
techniques, it is also possible to study the partial density of
states (PDOS). Hence, the need for a physically sound, minimal model
without any fictitious orbital and supported by accurate
parameterization is required to be able to provide realistic
descriptions of both valence and conduction bands in contrast to
simulating only the bandgap of these semiconductors.  Our
attempts~\cite{DD1,DDZnO} in this direction on the II-VI
semiconductors suggest that much of the difficulties arise from
inaccurate parameterization of the bulk band structures. In order to
overcome these difficulties, we carried out a detailed analysis of
the bulk band structure obtained within the highly accurate FP-LAPW
method, supported by the recently developed new generation
Muffin-Tin Orbital (NMTO) method~\cite{anderson} to obtain physical,
realistic and minimal model with accurate parameterizations. Such a
method not only obviates the need for the fictitious $s^*$ orbital,
but has also been found to explain quantum confinement effects quite
well in the II-VI semiconductors.~\cite{DDInP,DD2,DDZnO}

Though the synthesis of high quality nanometric sized II-VI
semiconductors is already very well established, the synthesis and
studies of high quality III-V semiconducting nanocrystals are being
increasingly reported in the recent
literature.~\cite{Brien,Gao,Guzelian,Malik,Manz}  III-V
semiconductors provide a material basis for a number of already
existing commercial products, as well as new cutting edge electronic
and optoelectronic devices, like heterostructure bipolar
transistors, diode lasers, light emitting diodes, electro-optic
modulators~\cite{optodevices} and in biology, as fluorescent
labels.~\cite{labels}  Hence, it becomes necessary to have an
electronic structure model with accurate predictive abilities to
describe the quantum confinement effects in these nanocrystals.

In order to achieve this, we study the band dispersions as well as
the DOS and PDOS obtained from the {\it ab-initio} Full Potential
Linearized Augmented Plane-Wave (FP-LAPW) method~\cite{LAPW} to
establish the relative importance of various orbital degrees of
freedom involved in describing the valence as well as some of the
low-lying conduction bands in various III-V semiconductors.  In
order to identify the dominant hopping interactions, we employ the
muffin-tin orbital (MTO)-based NMTO technique~\cite{anderson} which
provides an unique scheme to derive first-principles TB Hamiltonian
starting from full Local Density Approximation (LDA) calculation.
The usefulness of this method has been demonstrated in a number of
cases.~\cite{valenti}  With inputs from these {\it ab-initio}
methods, we construct a minimal TB model and carry out a
least-squared-error minimization procedure to fit the TB dispersions
to the {\it ab-initio} ones, thereby defining the values of the TB
parameters. Using the TB parameters thus obtained, we carry out a
real space calculation using Lanczos algorithm for different sizes
of the nanocrystals to obtain the dependence of the electronic
structure on the size of the nanocrystals. We compare these
calculated results with the experimentally determined bandgaps of
different nanocrystals as a function of size; the excellent
agreement in each case establishes the validity of the present
approach over the entire range of nanocrystal sizes. We have also
carried out a similar analysis on all the II-VI semiconductors
starting with the more accurate FP-LAPW results as inputs compared
to the previous approach.~\cite{DD1} As these results are found to
be slightly different and possibly more accurate in comparison to
the earlier results, the new parameter values for the II-VI series
are also reported here.

\section{Methodology:}

{\it Ab initio} band structures of all the III-V as well as II-VI
compound semiconductors with the zinc blende structure were obtained
using the FP-LAPW method.  Self consistency was achieved using 30
k-points in the irreducible Brillouin zone. In order to obtain a
realistic TB model, we first calculated the band dispersions and
density of states (DOS). The atomic orbital contributions to the
valence and conduction bands in the band dispersions and the DOS
were determined in terms of orbital-projected band-structure, the
so-called {\it fatbands} and the partial densities of states (PDOS),
respectively.  Analysis of these results establishes the minimal
orbital basis for the TB model.  To obtain a guideline for the range
of relevant hopping interactions necessary for reliable descriptions
of the valence and conduction bands, we carried out NMTO
calculations which provide TB Hamiltonians derived in a
first-principles way by constructing atom-centred, short-ranged
Wannier orbitals, namely the NMTOs.

The TB Hamiltonian is given by
\begin{equation}
{\bf H} = \sum_{il} \epsilon_l a^{\dag}_{il} a_{il} + \sum_{ij}
\sum_{ll'} (t^{ll'}_{ij} a^{\dag}_{il} a_{jl'}  + {\rm h.c.})
\end{equation}
where $a^{\dag}_{il}$ and $a_{il}$ are respectively the creation and
annihilation operators for electrons at the atomic site, $i$ in the
$l^{th}$ orbital. The onsite energy for the orbital $l$ at the site
$i$ is given by $\epsilon_{il}$.  The hopping interaction strengths
$t^{ll'}_{ij}$ depend on the type of orbitals and geometry of the
lattice and are parameterized using the Slater-Koster
parametrization scheme.~\cite{SlatKost}  We start with the estimates
of the values of onsite energies and the hopping integrals obtained
from NMTO derived TB Hamiltonian and then carry out  a least-squared
error minimization fitting procedure at a number of high symmetry
points in the band dispersion curves to fit the band dispersion
obtained from FP-LAPW method.  The parameters thus obtained are used
for calculations of the electronic structure of nanocrystals.

We generated the clusters consisting of a central anion surrounded
by the four nearest neighbor cations, followed progressively by
alternate shells of anion and cations, similar to our previous
studies on the different II-VI semiconductors,~\cite{DD2,DDZnO} and
on Mn doped GaAs.~\cite{DDGaAs} The effective diameter, $d$, of the
nanocrystal is calculated assuming that the particles are spherical
in shape using the formula
\begin{equation}
d = a [\frac{3N_{at}}{4{\pi}}]~^{1/3}
\end{equation}
where $a$ is the bulk lattice parameters and $N_{at}$ is the number
of atoms in the nanocrystal.  The largest nanocrystal for which the
DOS was calculated has $N_{at}$~$\sim$~10,000 atoms and
$d$~$\sim$~7.5~nm containing approximately 65,000 orbitals.  As it
is virtually impossible to perform a complete diagonalization of
such a large matrix, we obtain the eigen-value spectrum using the
Lanczos algorithm.~\cite{lanc}  We passivate the clusters with
hydrogen atoms at the outermost layer to remove the dangling bonds
and obtain the eigen-value spectrum for clusters with different
sizes.  From the eigen-spectrum, the top of the valence states (TVS)
and the bottom of the conduction states (BCS) are obtained and the
bare bandgap is calculated as the difference between them. However,
the bare bandgap cannot be directly obtained from experimental data,
usually based on optical absorption spectra, due to the presence of
the excitonic peak close to the absorption edge.~\cite{Liplan}
Hence, we compare the theoretically obtained excitonic peak position
with the experimental results. The excitonic peak position is
determined theoretically by subtracting the binding energy of the
exciton from the calculated bare bandgap. The excitonic binding
energy is given by the equation $E_c = 3.572e/\varepsilon d$, where
$\varepsilon$ is the dielectric constant of the material and $d$ is
the diameter of the nanocrystal.~\cite{brus}

\section{Results and Discussion:}

\subsection{TB parametrization of the bulk electronic structure:}
In order to obtain the physical and realistic model and accurate
parameter values for various semiconductors, we start with the
analysis of bulk band structure of these semiconductors.  Since the
successive steps involved in the analyses are similar for each of
the compounds, we illustrate the various steps using the example of
GaAs. The band dispersions obtained from the FP-LAPW method along
various symmetry directions for GaAs are shown in
Fig.~\ref{fatband}(a) with the zero of the energy scale referring to
the top of the valence band.  The calculated results show a direct
bandgap of about 0.3~eV, grossly underestimating the experimental
value of 1.4~eV. It is well known that LDA methods underestimate the
bandgap. However, since we are primarily interested in estimating
the change in the bandgap of a nanocrystal compared to that of the
bulk, we do not attempt to correct the bandgap artificially,
implicitly assuming that the errors in estimating the absolute
bandgaps cancel out to a large extent.  This assumption turns out to
be a reasonable one, as will be shown later in the text, for the
present series of III-V compounds and also for the II-VI
compounds.~\cite{DD2}

In order to understand the orbital contributions to the bulk band
structure, we obtain contribution of each of the orbitals to the
band wave-functions at each of the energy and momentum points, shown
in terms of fatbands, in Fig.~\ref{fatband}(b)-(f). In these panels,
though the band dispersions are same as in Fig.~\ref{fatband}(a),
only the fatness associated with each band varies, with the size of
the circles indicating the amount of the particular orbital
character for that band at that k-point. For example, it can be seen
that the lowest conduction band between 0 and 3~eV has contributions
mainly from Ga-$s$ (Fig.~\ref{fatband}(b)), with substantial mixing
from As-$s$ (Fig.~\ref{fatband}(d)) and As-$p$
(Fig.~\ref{fatband}(e)) states. However, main part of As-$s$ appears
as nearly flat band at an energy of $-$11~eV
(Fig.~\ref{fatband}(d)).  From Figs.~\ref{fatband}(b) and (e), it
can be seen that the three strongly dispersive valence bands between
$-$7~eV and 0~eV are made up of a mixture of mostly As-$p$ and
Ga-$s$ orbitals.  Though mostly prominent in the highly dispersive
bands above the lowest conduction band, Ga-$p$ contributions to some
parts of the lowest conduction band ({\it e.g.} along $\Gamma$-X and
near the X point) as well as to the top two valence bands are not
negligible (Fig.~\ref{fatband}(c)). Hence, it is necessary to
include at least the $sp^3$ orbitals of both Ga and As in the basis.
Also the contribution of As-$d$ (Fig.~\ref{fatband}(f)) to the bands
of interest, though not very prominent, cannot be neglected,
especially in the conduction band region.  Hence, we also included
As-$d$ in the basis. The contribution of Ga-$d$ (not shown in the
figure) however is virtually absent in the bands of interest,
suggesting a Ga $sp^3$- As $sp^3d^5$ as the suitable basis, in
contrast to the previously supported~\cite{Jancu}
$sp^3d^5s^*-sp^3d^5s^*$ basis.

While the atomic orbital contributions to the band wave-functions
along the high symmetry directions in terms of the fatbands provide
a clear suggestion for the suitable basis for the system, this can
be further supported by the analysis of the atomic orbital
contributions to the overall momentum averaged electronic structure
in terms of various PDOS shown in Fig.~\ref{dos}.  Panel (a), with
Ga-derived PDOS, shows that the bottom of the conduction band as
well as the sharp DOS feature at about $-$7~eV are dominated by
Ga-$s$ states.  Ga-$p$ states contribute significantly throughout
the conduction and valence states, while Ga-$d$ has very little
contribution in these energy ranges.  The lower panel (b) clearly
shows the dominance of the As-$p$ states in determining the valence
band states, while the conduction states have significant
contributions from As $s$, $p$ and $d$ states; in particular, we
find that As-$d$ states contribute nearly as much as the As- $s$
states in the conduction band region.

Having determined the relevant basis for the parameterized TB model,
we carried out a least-squared-error fit of the FP-LAPW band
dispersions in terms of the dispersions of this TB model by
systematically varying the electronic parameters (on-site and
hopping parameters) of the nearest neighbor (nn) TB Hamiltonian. The
fitting was carried out in two steps. First we carried out a fitting
of all the 13 bands arising primarily from the Ga-$s$ and $p$ and
the As-$s$, $p$ and $d$, though the As-$d$ states lie high in
energy. This inclusion of As-$d$ in the first step ensures that we
use a realistic value of As-$d$ for the fitting.  In the next step,
we fix the As-$d$ on-site energy and re-optimize the parameters to
fit the lowest 8 bands, in order to provide the most accurate
description of the relevant valence and conduction states, primarily
arising from the $s$ and $p$ orbitals of Ga and As.  The best fit
obtained this way is shown in Fig.~\ref{nn}(a).  From the figure, it
can be seen that though the basic features of the valence and the
low-lying conduction bands are captured in this approach, there are
too many important and gross discrepancies, such as, the curvature
of the lowest conduction band at the $\Gamma$ point, in the TB
results compared to the {\it ab-initio} results. We have highlighted
these discrepancies by boxes marked around such discrepancies in the
figure.

These significant discrepancies suggest that the model adopted here
misses out on some important interactions, thereby lacking the
desired level of accuracy.  Such problems have often prompted other
groups to increase the basis, for example by the {\it ad-hoc}
inclusion of an $s^*$ orbital on the cationic site.~\cite{Jancu}
However, this does not remedy the limitations of the model, as
confirmed by us by obtaining the best description TB dispersions
with Ga $sp^3s^*$- As $sp^3d^5$ basis in comparison to the {\it
ab-initio} approach; this comparison is shown in Fig.~\ref{nn}(b).

In order to obtain an insight into the possible origin of these
discrepancies, we carried out analysis based on NMTO calculations
that provide us a systematic and {\it ab-initio} way to construct
real space (RS) Hamiltonian by the Fourier transformation from the
usual momentum-space Hamiltonian.  The real space Hamiltonian,
generated in this manner, contains all different interactions,
ranging from the nearest-neighbor to the farthest interaction.
However it is possible to truncate RS Hamiltonian at various
distances, corresponding to different sized real spaced clusters and
back Fourier transform the truncated RS Hamiltonian to get the
corresponding tight-binding band dispersions. The tight binding
bands obtained from such a truncated Hamiltonian, when compared with
the band dispersions obtained from the complete calculation provide
an understanding of the important interactions present in a given
system.  The results of such analysis for GaAs are shown in
Fig.~\ref{nmto}. From panel (a), it can be observed that the
shortest ranged Hamiltonian including only the nearest neighbor
Ga-As interactions is not able to describe the conduction band
dispersions at all; one can also notice significant mismatches
within the valence band region as well.  An extension of the range
of the Hamiltonian to additionally include As-As next nearest
neighbor interactions (panel (b)) improves significantly the
description of the conduction band states, thereby establishing the
importance of As-As interactions in determining the electronic
structure of this compound.  However we still find substantial
discrepancies and hence include the Ga-Ga interaction.  In this
case, we find a substantial overall improvement in the descriptions
of the valence band dispersions as well as the conduction bands, as
shown in panel (c).  This suggests that the most reasonable
parameterized TB Hamiltonian with the Ga $sp^3$ - As $sp^3d^5$ basis
should include the nearest neighbor and the second nearest neighbor
interactions, where suitably chosen interaction parameters will be
able to provide a satisfactory description to the electronic
structure of GaAs via renormalization of these parameters to include
effects of all those interactions that are neglected in this minimal
basis, short ranged TB model.

The above-mentioned expectation is comprehensively justified by the
results shown in Fig.~\ref{2nn}, where we present the fits to the
{\it ab-initio} band dispersions within three different
parameterized TB models, namely (a) Ga $sp^3$-As $sp^3d^5$ basis
with Ga-As and As-As interactions; (b) Ga $sp^3s^*$-As $sp^3d^5$
basis with Ga-As and As-As interactions; and (c) Ga $sp^3$ - As
$sp^3d^5$ basis with Ga-As, Ga-Ga and As-As interactions.
Fig.~\ref{2nn}(c) evidently exhibits the most accurate description
of the electronic structure of GaAs over the entire valence and
conduction band ranges in terms of the TB model.  These results
further establish that it is not necessary to introduce the
fictitious $s^*$ orbital in the basis, as it does not improve
anything significantly.

Similar analysis were carried out for all the other III-V systems
studied here.  In every case, except for GaN we found the model with
the cationic $sp^3$- anionic $sp^3d^5$ basis and first and second
nearest neighbor interactions to be both necessary and sufficient to
provide accurate descriptions of the electronic structures.  In the
case of GaN, Ga $sp^3d^5$ - N $sp^3$ basis was found to be the most
suitable.  The comparison between the {\it ab-initio} band
dispersion and the TB dispersion with the optimized electronic
parameter strengths is shown for each of the compounds in
Fig.~\ref{allfit}, illustrating highly accurate descriptions
throughout.  The corresponding optimized parameter values are given
in Table~\ref{para}.

An earlier study of II-VI compounds employing band dispersions
calculated within the Linearized Muffin-Tin Orbitals (LMTO) and
Atomic sphere approximation (LMTO-ASA) as the reference for the
electronic structure provided~\cite{DD1} a TB model based on the
$sp^3d^5$ basis on both cations and anions and the nearest neighbor
cation-anion and second nearest neighbor anion-anion interactions
only.  Noting that FP-LAPW provides a more accurate starting point
compared to the LMTO-ASA results and that the earlier model for
II-VI compound is slightly different from the present model for the
III-V compounds, we have reinvestigated the II-VI series employing
FP-LAPW calculations.  Carrying out a similarly detailed analysis as
presented here for the III-V compounds, we found that the  minimal
basis for accurate descriptions of electronic structures for the
II-VI series consists of the previously employed $sp^3d^5$ orbitals
on both anions and cations and both the second nearest neighbor
interactions in addition to the nearest neighbor cation-anion
interactions.  In essence, the present analyses suggest that TB
model of Ref. [9] needs to be extended to include also the second
nearest neighbor cation-cation interaction in order to provide a
comparable level of accuracy in describing the FP-LAPW results.
While we do not present the details of the analysis for the II-VI
series here, it being along the same line as presented for the III-V
series, we have tabulated the optimized electronic parameter
strengths of the TB model in Table~\ref{paraii}.

\subsection{Electronic structure of nanocrystals:}
The bandgap of a finite sized crystal is known to have a pronounced
dependence on the size of the crystal in the nanometric regime. This
has opened up immense technological possibilities, based primarily
on the tunability of the bandgap in the quantum confinement regime.
If the electronic parameter strengths as well as the TB model itself
remain valid down to such small sizes, the present analysis and
results provide a reliable way to understand or even predict such
bandgap variations with the size by performing real space
calculations for the finite sized crystals with the same model and
parameter values.  Encouraged by the previous success in similar
studies,~\cite{DD1,DD2,DDGaAs,DDInP,DDZnO} we have carried out
electronic structure calculations for finite sized III-V systems
using the Lanczos algorithm as described in Section II.  Various
panels in Fig.~\ref{variation} show the calculated shifts (open
circles) in the bandgap of nanocrystal relative to the bulk bandgap
as a function of size.  In order to provide an analytical
description of the systematic variation of $\Delta E_g$, the
calculated results (open circles) were fitted with an empirical
expression of the form $1/(ad^2 + bd + c)$ where $a$, $b$ and $c$
are obtained by fitting.  The choice of the expression here, though
entirely empirical, was prompted by the $1/d^2$ dependence found in
EMA.  However, this simple dependence was found to be insufficient
to describe the results; therefore we use the simplest extension of
the EMA expectation that is able to fit the results accurately
enough.  The values of $a$, $b$ and $c$ for different semiconductors
are shown in Table~\ref{fitpara}; these values allow one to
calculate the change in bandgap of a nanocrystal with any specific
size.  The curve obtained by fitting is shown in the various panels
of Fig.~\ref{variation} as a solid thick line. In this figure, we
also compare this calculated curve with experimental data, wherever
available, different symbols representing data from different
publications. We have also compiled in these figures calculated
results from other approaches, such as those based on EMA (dashed
lines in each panel), TB model using $sp^3d^5s^*$ basis~\cite{Allan}
(dotted lines) and semi-empirical pseudo-potential
method~\cite{ppmeth2} available only for InP (dashed dotted line).
These comparisons clearly show that the present TB model provides a
description of the experimentally observed variation of bandgaps
more accurately than the other theoretical approaches.  For example,
EMA is found to grossly overestimate the bandgaps in every case. The
$sp^3d^5s^*$ model is also found to overestimate the bandgap
variation compared to the experimental data for InAs (see panel (i)
of Fig.~\ref{allfit}); in contrast, results from the present model
is found to be in striking agreement with the experimental results.
For InP, the only other case where extensive experimental results
exist, we again find a remarkable agreement with calculated results
based on the present model over the entire range of sizes.  This
establishes the effectiveness of the TB model developed here and
reliability of the estimated parameter strengths even for the study
of finite sized nanocrystals.

As we have a new set of parameter values (Table~\ref{paraii}) with a
slightly different TB model for the II-VI series compared to the
earlier report,~\cite{DD1} we have carried the electronic structure
calculations for the nanocrystals of all these II-VI compounds also,
for the sake of completion.  We find that the new results are in
good agreement with the experimental data reported in Ref. [11]
earlier. The variation of bandgap in the II-VI semiconductors are
also fitted using the same expression $\Delta E_g = 1/ (ad^2 + bd +
c)$ and the values of the fitting parameters are shown in
Table~\ref{fitparaii}.

In view of the recent experiments on II-VI semiconductors using high
energy spectroscopies, mapping out separately the valence and
conduction bands,~\cite{ddjag} it is important to understand the
variations of TVS and BCS separately, in addition to probing the
changes in the bandgap with size.   Since the calculated bandgap is
constructed from the difference in TVS and BCS, it is straight
forward to calculate the variation of the TVS and BCS from our
calculations.  The variations of TVS (open circles) and BCS (closed
circles) as a function size for the various III-V semiconductors are
shown in different panels of Fig.~\ref{tvbbcb}. Since we show the
change in these quantities as a function of the size with respect to
those for the bulk, the zero of the energy axis corresponds to the
bulk values; it can be seen from the figures that both TVS and BCS
smoothly approach the bulk values with increasing size of the
nanocrystals.  We also observe that the shift in BCS is larger than
that of the TVS in the larger size regime for most of the systems;
this indicates that the shift in the bandgap is dominated by the
shift in the conduction band edge in such cases.  The predominance
of the BCS in determining the variation of the bandgap is easy to
understand in terms of effective masses of electrons and holes.
First, we note that the energy variation of electron or hole states
are related inversely to the corresponding effective masses; in
other words, the shifts in the valence and conduction states are
controlled by 1/$m_h^*$ and 1/$m_e^*$, respectively.  Since the
$m_e^*$ is significantly larger than $m_h^*$ for the III-V
compounds, the conduction band is affected more pronouncedly
compared to the valence band with a change in the size.  This
argument is valid only for the larger size limit where effective
mass, determined for the bulk material remains to be a relevant
quantity even for the nanocrystals.  However, in the smaller size
regime the trend appears to be reversed in several cases, such as in
GaAs (Fig.~\ref{tvbbcb} (f)), with the shift in the BCS being less
than that in the TVS.  This change in the behavior with size can be
understood in the following way.  At large sizes, the BCS is defined
by the states belonging to the $\Gamma$ point of the bulk band
structure (see Fig.~\ref{allfit} (f) for GaAs) which has a low
effective electron mass, being dominantly contributed by the Ga-$s$
states.  However, a rapid upward movement of these states with
decreasing size inevitably makes BCS to be contributed primarily by
the states belonging to X point of the bulk band structure (see
Fig.~\ref{allfit}(f)) which evidently corresponds to a larger
effective mass, being primarily contributed by the Ga-$p$ states and
having a relatively flat dispersion.  Therefore, this leads to a
relatively less pronounced change in the BCS in the smaller size
regime.

\section{Conclusions:}
In this work, we present a systematic development of parameterized
tight-binding model for an accurate description of the electronic
structure of group III-V semiconductors. We analyze the nature and
origin of bonding as well as the atomic orbital contributions to
each band eigen-states to arrive at the necessary minimal model
involving $sp^3$ orbitals at the cationic sites and $sp^3d^5$
orbitals at the anionic sites, obviating the use of any fictitious
$s^*$ orbital in the basis. We find that though the
nearest-neighbor-only model provides an approximate description of
the {\it ab initio} band dispersions over a wide energy range, it is
necessary to include both the cation-cation and anion-anion second
nearest neighbor interactions to obtain a satisfactorily accurate
description of {\it ab-initio} band dispersions. We have also
performed similar analysis for the II-VI semiconductors using the
more accurate FP-LAPW {\it ab-initio} band structure in contrast to
the previously used LMTO method. Using these optimized parameters,
we perform real space calculations with the same tight binding model
to obtain the variation of the bandgap as a function of the
nanocrystal size. A comparison with the available experimental data
of the bandgap variation with the size of these nanocrystals
exhibits good agreement over the entire range of sizes; in sharp
contrast to the results obtained with the EMA. We have also compared
the present results with other calculations and we find that the
present results give a better description, wherever these differ. A
similar analysis was also carried out on the II-VI semiconductors
using the newly obtained parameters and the calculated bandgap
variation is are found to match well with existing experimental
values.  Ideally one would like to extend similar parameterized
tight-binding Hamiltonian approach, not only for an accurate
description of the electronic structure of such systems, but also to
describe the cohesive energy and geometry optimization; this will,
however, require an accurate description of the ionic contributions
of the total energy along with the electronic contributions that has
been modeled here.

{\bf Acknowledgements:} This work is supported by the Department of
Science and Technology, Government of India.  We thank P. Blaha, K.
Schwarz, P. Dufek and R. Augustyn for providing the LAPW code.  We
thank Dr. Aparna Chakrabarti for helpful discussions.

\newpage

\begin{table}
\centering \caption{\label{para} Parameters (in eV) obtained from TB
least-squared-error fitting procedure for the various III-V
semiconductors.  The parameters are obtained with energy zero at valence band maximum.}
\begin{centering}
{
\begin{tabular}{rrrrrrrrrrr}
\hline
&~~~~~~AlP&~~~~~~AlAs&~~~~~~AlSb&~~~~~~GaN&~~~~~~GaP&~~~~~~GaAs&~~~~~~Gasb&~~~~~~InP&~~~~~~InAs&\\
\hline
$s_c$ & 4.89 & 4.22 & 4.37 & 6.02 & 1.73 & 1.12 & 1.01 & 1.50 & 1.04 &\\
$p_c$ & 8.44 & 8.24 & 5.71 & 10.60 & 7.45 & 7.79 & 6.38 & 7.00 & 5.76 &\\
$s_a$ & $-$8.24 & $-$9.12 & $-$8.26 & $-$9.58 & $-$8.48 & $-$10.15 & $-$9.50 & $-$8.00 & $-$9.74 &\\
$p_a$ & $-$0.62& $-$0.14 & 0.69 & $-$0.51 & $-$0.04 & $-$0.31 & $-$1.04 & 0.53 & 0.06 &\\
$d_a$ & 9.43 & 9.31 & 6.80 & $-$ & 7.87 & 6.80 & 7.25 & 9.00 & 8.25 &\\
$s_cs_a \sigma$ & $-$1.39 & $-$1.10 & $-$1.26 & $-$0.53 & $-$1.78 & $-$1.26 & $-$0.47 & $-$1.04 & $-$0.82 &\\
$s_cp_a \sigma$ & 2.42 & 2.41 & 2.67 & 1.76 & 2.81 & 2.78 & 2.28 & 2.30 & 2.34 &\\
$s_cd_a \sigma$ & $-$1.65 & $-$1.76 & $-$2.63 & $-$ & $-$2.27 & $-$2.00 & $-$1.93 & $-$1.72 & $-$1.81 &\\
$p_cp_a \sigma$ & 2.79 & 2.42 & 3.01 & 3.66 & 3.55 & 2.86 & 2.23 & 2.99 & 2.88 &\\
$p_cp_a \pi$ & $-$0.57 & $-$0.79 & $-$0.70 & $-$1.12 & $-$0.83 & $-$1.04 & $-$0.74 & $-$0.53 & $-$0.63 &\\
$p_cd_a \sigma$ & $-$0.09 & $-$1.33 & $-$1.71 & $-$ & $-$1.60 & $-$0.85 & $-$1.02 & $-$0.05 & $-$0.04 &\\
$p_cd_a \pi$ & 2.47 & 1.71 & 1.43 & $-$ & 1.80 & 1.39 & 1.45 & 2.12 & 2.00 &\\
$p_cs_a \sigma$ & $-$1.87 & $-$1.47 & $-$1.21 & $-$4.50 & $-$1.81 & $-$0.48 & $-$0.22 & $-$1.63 & $-$1.19 &\\
$s_cs_c \sigma$ & $-$0.23 & $-$0.28 & $-$0.46 & $-$0.30 & $-$0.36 & $-$0.27 & $-$0.12 & $-$0.18 & $-$0.21 &\\
$s_cp_c \sigma$ & 0.23 & 0.34 & 0.38 & 0.01 & 0.11 & 0.20 & 0.03 & 0.15 & 0.05 &\\
$p_cp_c \sigma$ & 0.35 & 0.57 & 0.35 & 1.39 & 0.48 & 0.13 & 0.32 & 0.00 & 0.05 &\\
$p_cp_c \pi$ & $-$0.34 & $-$0.15 & $-$0.14 & $-$0.57 & $-$0.23 & $-$0.01 & $-$0.22 & $-$0.20 & $-$0.23 &\\
$s_as_a \sigma$ & $-$0.08 & $-$0.08 & $-$0.01 & $-$3.24 & $-$0.00 & $-$0.01 & $-$0.15 & $-$0.12 & $-$0.10 &\\
$s_ap_a \sigma$ & 0.25 & 0.05 & 0.01 & 1.31 & 0.14 & 0.10 & 0.22 & 0.12 & 0.16 &\\
$s_ad_a \sigma$ & $-$0.05 & 0.00 & $-$0.08 & $-$ & $-$0.07 & $-$0.16 & $-$0.39 & $-$0.16 & $-$0.11 &\\
$p_ap_a \sigma$ & 0.49 & 0.28 & 0.32 & 1.21 & 0.46 & 0.17 & 0.35 & 0.47 & 0.41 &\\
$p_ap_a \pi$ & $-$0.04 & $-$0.04 & $-$0.01 & 0.02& $-$0.02 & $-$0.03 & $-$0.01 & $-$0.00 & $-$0.01 &\\
$p_ad_a \sigma$ & $-$0.13 & $-$0.33 & $-$0.22 & $-$ & $-$0.30 & $-$0.32 & $-$0.32 & $-$0.02 & $-$0.00 &\\
$p_ad_a \pi$ & 0.00 & 0.08 & 0.00 & $-$ & 0.05 & 0.10 & 0.11 & 0.03 & 0.01 &\\
$d_ad_a \sigma$ & $-$1.16 & $-$0.76 & $-$0.85 & $-$ & $-$1.02 & $-$0.85 & $-$0.75 & $-$1.06 & $-$0.85 &\\
$d_ad_a \pi$ & 0.33 & 0.50 & 0.37 & $-$ & 0.42 & 0.49 & 0.39 & 0.34 & 0.27 &\\
$d_ad_a \delta$ & $-$0.01 & $-$0.01 & $-$0.01 & $-$ & $-$0.01 & $-$0.14 & $-$0.01 & $-$0.00 & $-$0.00 &\\

\hline
\end{tabular}

\begin{tabular}{rrrrrrrrrrrr}
~~~~ &~~~~~ $d_c$ & ~~~~~$d_cs_a\sigma$ & ~~~~$d_cp_a\sigma$ & ~~~~$d_cp_a\pi$ & ~~~~$s_cd_c\sigma$ & ~~~~$p_cd_c\sigma$ & ~~~~$p_cd_c\pi$ & ~~~~$d_cd_c\sigma$ & ~~~~$d_cd_c\pi$ & ~~~~$d_cd_c\delta$ &\\
\hline
GaN & 13.51 & $-$1.13 & 2.13 & $-$1.81 & $-$0.66 & $-$1.89 & 0.34 & $-$2.38 & 0.55 & 0.00 &\\
\hline
\end{tabular}

}
\end{centering}
\end{table}

\begin{table}
\centering \caption{\label{paraii} Parameters (in eV) obtained from
TB least-squared-error fitting procedure for the various II-VI
semiconductors.  The parameters are obtained with energy zero at
valence band maximum.}
\begin{centering}
{

\begin{tabular}{rrrrrrrr}
\hline
&~~~~~~ZnS&~~~~~~ZnSe&~~~~~~ZnTe&~~~~~~CdS&~~~~~~CdSe&~~~~~~CdTe&\\
\hline
$s_c$ & 1.18 & $-$0.33 & 3.26 & 3.53 & 2.66 & 2.50 & \\
$p_c$ & 9.79 & 6.60 & 5.05 & 8.46 & 8.28 & 7.16 & \\
$d_c$ & -6.46 & $-$6.70 & $-$6.86 & $-$7.53 & $-$7.56 & $-$7.96 & \\
$s_a$ & $-$8.68 & $-$9.21 & $-$10.05 & $-$10.88 & $-$10.75 & $-$9.68 & \\
$p_a$ & 0.47 & 1.35 & $-$0.22 & $-$0.41 & $-$0.49 & $-$0.62 & \\
$d_a$ & 11.27 & 10.91 & 9.45 & 13.60 & 10.24 & 9.12 & \\
$s_cs_a \sigma$ & $-$1.65 &  $-$1.61 & $-$1.03 & $-$1.02 & $-$0.94 & $-$0.72 & \\
$s_cp_a \sigma$ & 2.41 & 2.41 & 1.36 & 1.83 & 1.90 & 1.77 & \\
$s_cd_a \sigma$ & $-$1.71 & $-$1.82 & $-$1.88 & $-$2.41 & $-$1.81 & $-$1.90 & \\
$p_cp_a \sigma$ & 3.73 & 3.67 & 3.05 & 2.75 & 2.73 & 2.34 & \\
$p_cp_a \pi$ & $-$0.46 & $-$0.61 & $-$0.53 & $-$0.30 & $-$0.39 & $-$0.38 & \\
$p_cd_a \sigma$ & $-$0.32 & $-$0.92 & $-$0.95 & $-$0.56 & $-$0.13 & $-$0.33 & \\
$p_cd_a \pi$ & 3.21 & 2.49 & 1.41 & 2.53 & 2.47 & 2.21 & \\
$p_cs_a \sigma$ & $-$2.31 & $-$1.86 & $-$1.15 & $-$1.07 & $-$2.13 & $-$1.59 & \\
$d_cs_a \sigma$ & $-$1.42 & $-$1.49 & $-$0.52 & $-$0.68 & $-$0.65 & $-$0.52 & \\
$d_cp_a \sigma$ & 0.01 & 0.01 & 0.01 & 0.97 & 0.95 & 0.82 & \\
$d_cp_a \pi$ & $-$0.13 & $-$0.10 & $-$0.06 & $-$0.40 & $-$0.34 & $-$0.25 & \\
$s_cs_c \sigma$ & $-$0.24 & $-$0.19 & $-$0.30 & $-$0.33 & $-$0.29 & $-$0.23 & \\
$s_cp_c \sigma$ & 0.11 & 0.00 & 0.01 & 0.07 & 0.20 & 0.16 & \\
$p_cp_c \sigma$ & 0.01 & 0.01 & 0.29 & 0.31 & 0.34 & 0.14 & \\
$p_cp_c \pi$ & $-$0.24 & $-$0.01 & $-$0.24 & $-$0.20 & $-$0.22 & $-$0.15 & \\
$s_as_a \sigma$& $-$0.05 & $-$0.02 & 0.04 & 0.00 & $-$0.07 & $-$0.07 & \\
$s_ap_a \sigma$ & 0.19 & 0.19 & 0.01 & 0.13 & 0.01 & 0.01 & \\
$p_ap_a \sigma$ & 0.55 & 0.53 & 0.78 & 0.33 & 0.32 & 0.35 & \\
$p_ap_a \pi$ & $-$0.01 & $-$0.09 & $-$0.10 & 0.00 & $-$0.02 & $-$0.02 & \\
\hline
\end{tabular}

}
\end{centering}
\end{table}

\begin{table}
\centering \caption{\label{fitpara} Parameters obtained from fitting
the variation of bandgap for the different III-V semiconductors
using the form $\Delta E_g = 1/(ad^2 + bd + c)$.}
{
\begin{tabular}{|ccccc|}
\hline
&~$a$ (nm$^{-2}$eV$^{-1}$)~&~$b$ (nm$^{-1}$eV$^{-1}$)~&~$c$ (eV$^{-1}$)~&\\
\hline
AlP & 0.1605 & $-$0.0588 & 0.2663 &\\
AlAs & 0.0997 & 0.1477 & 0.0279 & \\
AlSb & 0.1258 & $-$0.0649 & 0.2072 & \\
GaN & 0.3716 & $-$0.2336 & 0.2172 & \\
GaP & 0.1969 & 0.2631 & 0.0728 & \\
GaAs & 0.0359 & 0.1569 & 0.1564 & \\
GaSb & 0.0357 & 0.1963 & 0.1175 & \\
InP & 0.0461 & 0.3153 & 0.0623 & \\
InAs & 0.0374 & 0.2569 & 0.1009 & \\

\hline
\end{tabular}
}
\end{table}

\begin{table}
\centering \caption{\label{fitparaii} Parameters obtained from
fitting the variation of bandgap for the different II-VI
semiconductors using the form $\Delta E_g = 1/(ad^2 + bd + c)$.}
{
\begin{tabular}{|ccccc|}
\hline
&~$a$ (nm$^{-2}$eV$^{-1}$)~&~$b$ (nm$^{-1}$eV$^{-1}$)~&~$c$ (eV$^{-1}$)~&\\
\hline
ZnS & 0.2349 & $-$0.0418 & 0.2562 &\\
ZnSe & 0.0845 & 0.1534 & 0.2128 & \\
ZnTe & 0.0092 & 0.1872 & 0.2396 & \\
CdS & 0.1278 & 0.1018 & 0.1821 & \\
CdSe & 0.0397 & 0.1723 & 0.1111 & \\
CdTe & 0.0275 & 0.2403 & 0.1469 & \\

\hline
\end{tabular}
}
\end{table}

\clearpage

\begin{figure}
\caption{\label{fatband} (a) FP-LAPW band dispersions for the zinc
blende structure of GaAs along the various symmetry lines. (b)-(f)
Fatbands showing respectively the contribution of Ga-$s$, Ga-$p$,
As-$s$, As-$p$ and As-$d$ on various bands.}
\end{figure}

\begin{figure}
\caption{\label{dos}  (color online) PDOS corresponding to atomic
various orbitals in GaAs.  In Fig.~\ref{dos}(a), the thin solid
line, dotted line and the dashed line represents PDOS of Ga-$s$,
Ga-$p$ and Ga-$d$ orbitals respectively. In Fig.~\ref{dos}(b), the
thin solid line, dotted line and dashed line represents the PDOS of
As-$s$, As-$p$ and As-$d$ respectively.}
\end{figure}

\begin{figure}
\caption{\label{nn} (color online) Comparisons of band dispersions
obtained for zinc-blende structure of GaAs, from FP-LAPW and from TB
fitting for the nearest-neighbor interactions only in the (a)
$sp^3-sp^3d^5$ -orbital basis and (b) $sp^3s^*-sp^3d^5$ -orbital
basis on Ga and As respectively.  The open circles represent the
FP-LAPW calculation and the solid line represents the TB
calculation.}
\end{figure}

\begin{figure}
\caption{\label{nmto}  (color online) Comparison of TB bands of
various hopping ranges computed within the NMTO scheme with the LDA
band-structure. The LDA band-structure is shown as open circles,
while the TB bands are shown as thick solid line. The hopping ranges
include (a) nearest neighbor interaction (b) nearest neighbor and
As-As interaction (c) nearest neighbor and Ga-Ga as well as As-As
interactions in the $sp^3-sp^3d^5$ -orbital basis on Ga and As
respectively.}
\end{figure}

\begin{figure}
\caption{\label{2nn} (color online) Comparisons of band dispersions
obtained from FP-LAPW and TB error minimized fit obtained using
nearest neighbor and (a) As-As interactions in the $sp^3-sp^3d^5$ -
orbital basis, (b) As-As interactions in the $sp^3s^*-sp^3d^5$ -
orbital basis and (c) Ga-Ga as well as As-As interactions in the
$sp^3-sp^3d^5$ - orbital basis on Ga and As respectively.  The open
circles represent the FP-LAPW calculation and the solid line
represents the TB calculation.}
\end{figure}

\begin{figure}
\caption{\label{allfit} (color online) Band dispersions of various
III-V semiconductors obtained using FP-LAPW (open circles) and TB
fit (solid line) obtained using the optimized parameters, given in
Table~\ref{para}.}
\end{figure}

\begin{figure}
\caption{\label{variation} (color online) Variation of bandgap of
the different III-V nanocrystals obtained from TB approximation and
comparison with experimentally obtained data (panel (d)
refs~\cite{grocholl} (closed circles),~\cite{micic2} (open
circles),~\cite{cao} (open triangle), panel (e) refs~\cite{Gao}
(open circle),~\cite{micic1} (open triangle), panel (f)
refs~\cite{Olshavsky} (open circles),~\cite{Uchida} ( open
triangles),~\cite{malik} (closed circles),~\cite{Kher} (closed
triangles), panel (g) ref~\cite{lui} (open circles), panel (h)
refs~\cite{micic3} (open circles),~\cite{micic4} (closed
circles),~\cite{micic5} (closed triangles), panel (i)
refs~\cite{Guzelian} (open circles),~\cite{banin} (open
triangles),~\cite{talapin} (closed circles),~\cite{labels} (closed
triangles) ). The curves obtained from EMA are shown by the dashed
line and the $sp^3d^5s^*$ calculation are shown by dashed dotted
lines. }
\end{figure}

\begin{figure}
\caption{\label{tvbbcb} Difference in TVS (open circles) and BCS
(closed circles) from the bulk value plotted as a function of the
nanocrystal size for the various III-V semiconductors. The solid
line is a guide to eye connecting the data smoothly.}
\end{figure}

\end {document}